\documentclass[aps,twocolumn,pra,showpacs]{revtex4}
\usepackage{graphicx}
\usepackage{amsmath}
\usepackage{amssymb}
\usepackage{amsthm}
\usepackage{amsbsy}
\usepackage{epsfig}
\usepackage{dcolumn}

\newcommand{\beq}{\begin{equation}}
\newcommand{\eeq}{\end{equation}}
\newcommand{\bea}{\begin{eqnarray}}
\newcommand{\eea}{\end{eqnarray}}

\makeatletter
\def\btt#1{\texttt{\@backslashchar#1}}
\DeclareRobustCommand\bblash{\btt{\@backslashchar}}
\makeatother

\begin{document}

\author{Zhao-Ming Wang\thanks{%
Email address: mingmoon78@126.com},$^{1,2}$ Mark S. Byrd,$^{1,3}$\thanks{%
Email address: mbyrd@physics.siu.edu}, Bin Shao $^{2}$ and Jian Zou $^{2}$}
\affiliation{$^{1}$ Department of Physics, Southern Illinois University, Carbondale,
Illinois 62901-4401}
\affiliation{$^{2}$ Department of Physics, Beijing Institute of Technology, Beijing,
100081, China}
\affiliation{$^{3}$ Department of Computer Science, Southern Illinois University,
Carbondale, Illinois 62901-4401}
\title{Robust and Reliable Transfer of a Quantum State Through a Spin Chain}

\begin{abstract}
We present several protocols for reliable quantum state transfer through a spin chain. We use a simple two-spin encoding to achieve a remarkably high
fidelity transfer for an arbitrary quantum state. The fidelity of the transfer also decreases very slowly with increasing chain length. We find that we
can also increase the reliability by taking advantage of a local memory and/or confirm transfer using a second spin-chain.
\end{abstract}

\maketitle











\section{Introduction}

One of (Di Vincenzo's) seven requirements of a quantum computer is the transmission of flying qubits, meaning, we must be able to transmit information
between components of a quantum computing device. In the interior of a such a device, where short-distance communication is required, a spin chain is a
promising candidate for the transfer of information.

 Reliable transfer through a spin chain has been studied extensively since the original proposal by
Bose \cite{Bose:03}. In that proposal, a spin state at one end of the chain is allowed to evolve freely under constant couplings until it arrives after
some time at the other end of the chain. Such a system is simple and does not require couplings to be precisely tuned or to be switched on and off. This
is desirable for experimental considerations where controllability can be a definite problem.

It is indeed somewhat surprising that a spin state can be reliably
transferred through a chain. However, it certainly can be done when
particular conditions are met. For example, with finely-tuned, yet fixed
couplings, a variety of networks will allow for perfect transfer \cite%
{Christandl/etal:04,Christandl/etal:05}. There are also methods
using \textsl{two} chains which allow for perfect transfer \cite%
{Burgarth/Bose:05,Burgarth/etal:06}. In this case the perfect transfer is
conditioned on the outcome of two measurements, one from each chain. Other
methods require a wave packet to be constructed at the beginning of the
chain so that the state can be transferred reliably \cite%
{Osborne/Linden:04,Haselgrove:05}.

In each of these scenarios, the state is transferred using an always-on
Heisenberg exchange interaction with nearest-neighbor interactions between
the spins. The Heisenberg exchange interaction is readily available in many
different experimental systems. Thus its use is well-motivated. However, its
experimental viability is important for another reason--it enables universal
quantum computation on decoherence-free subspaces and noiseless subsystems
without the need for individual control over physical qubits. (See \cite%
{Byrd/etal:pqe04} and references therein.) There are several documented instances of this. One promising proposal uses two spins, or qubits, to encode
one logical qubit \cite{Palma:96}. However, it is also known to be universal for DFS/NSs using three or four physical qubits to encode one logical
qubit. In our case, the universality condition, prompted the following question, ``Can a decoherence-free, or noiseless, encoding be used to enable the
reliable propagation of a spin state through a spin chain.'' In this article, we answer this question and show that a particular state can be used to
reliably transfer quantum information over long distances through a spin chain. Our proposal uses encoded states which provide reliable state transfer
over relatively long distances through an unmodulated spin chain. This makes our proposal a prime candidate for use in experimental systems where this
sort of state transfer is required. Namely within a solid-state quantum computing device.


\section{The Hamiltonian and the calculation of the fidelity}

The Hamiltonian of a one dimension anisotropic Heisenberg XY model can be
described by 
\begin{equation*}
H=-\frac{J}{2}\sum\limits_{i=1}^{N-1}(\sigma _{i}^{x}\sigma
_{i+1}^{x}+\sigma _{i}^{y}\sigma
_{i+1}^{y})+J_{z}\sum\limits_{i=1}^{N-1}\sigma _{i}^{z}\sigma
_{i+1}^{z}-h\sum\limits_{i=1}^{N}\sigma _{i}^{z}.
\end{equation*}
where $J$ is the exchange constant for the $xy$ components, $J_{z}$ is the exchange constant for the $z$ component, and $h$ is external magnetic field
along $z$ direction. The $\sigma _{i}^{x,y,z}$ are the Pauli operators acting on the $i^{th}$ spin. We will use a ferromagnetic coupling and take
$J=1.0$ throughout this paper. Furthermore, we consider only nearest-neighbor interactions and an open ended chain which is the most natural and
practical geometry for this system.

Note that for this Hamiltonian, the $z$-component of the total spin, $%
\sigma^{z}=\sum\sigma_{i}^{z}$, is a conserved quantity, which indicates
that the system contains a fixed number of magnon excitations. When there is
only one magnon excitation, the time evolution of the initial state is not
affected by the $\sigma^{z}-\sigma^{z}$ interaction, whereas for the
two-magnon excitations it is. This can be quite complicated \cite%
{Subrahmanyam:04} due to the magnon interactions arising from a nonzero $%
J_{z}$. For these reasons, we let $J\gg J_{z}$.

The Hamiltonian can be diagonalized by means of the Jordan-Wigner
transformation that maps spins to one-dimensional spinless fermions with
creation operator defined by $c_{l}^{\dag }=(\prod\limits_{s=1}^{l-1}-\sigma
_{l}^{z})\sigma _{l}^{+}$, where $\sigma _{l}^{+}=\frac{1}{2}$ $(\sigma
_{l}^{x}+i\sigma _{l}^{y})$ denotes the spin raising operator at site $l$.
The action of $c_{l}^{\dag }$ is to flip the spin at site $l$ from down to
up and $c_{l},c_{m}^{\dag }$ satisfy the anticommutation relations $%
\{c_{l},c_{m}^{\dag }\}=\delta _{lm}$.

The creation operator evolves as \cite{Amico/etal:04b}
\begin{equation}  \label{eq:oe}
c_{j}^{\dag }(t)=\sum_{l=1}^{N}f_{j,l}(t)c_{l}^{\dag }, \;
\end{equation}
where $N$ is the number of spins,
\begin{equation}
f_{j,l}(t)=\frac{2}{N+1}\sum\limits_{m=1}^{N}\sin (q_{m}j)\sin
(q_{m}l)e^{-iE_{m}t},
\end{equation}
$E_{m}=2h-2J\cos q_{m}$, and $q_{m}=\pi m/(N+1)$. Eq.~(\ref{eq:oe})
indicates that the excitation which, initially created in site $j$, is
generally distributed over all the sites. At time $t$, the probability of
the excitation being at site $l$ is $\left\vert f_{j,l}(t)\right\vert ^{2}$
with the normalization condition $\sum_{l=1}^{N}\left\vert
f_{j,l}(t)\right\vert ^{2}=1$.

When the number of magnon excitations is more than one, the time evolution
of the creation operators is given by \cite{Wichterich/Bose:08}
\begin{equation}
\prod\limits_{m=1}^{M}c_{j_{m}}^{\dag }(t)=\sum\limits_{l_{1},..,l_{M}}\det
\left\vert
\begin{array}{cccc}
f_{j_{1},l_{1}} & f_{j_{1},l_{2}} & ... & f_{j_{1},l_{M}} \\
f_{j_{2},l_{1}} & f_{j_{2},l_{2}} & ... & f_{j_{2},l_{M}} \\
... & ... & ... & ... \\
f_{j_{M},l_{1}} & f_{j_{M},l_{2}} & ... & f_{j_{M},l_{M}}%
\end{array}%
\right\vert \prod\limits_{m=1}^{M}c_{l_{m}}^{\dag }.  \label{eq:ffe}
\end{equation}
where $M$ is the number of the excitations. The row \{$j_{1},j_{2},...,j_{M}$%
\} denotes the sites where the excitations are created and \{$%
l_{1},l_{2},...,l_{M}$\} ($l_{1}<l_{2}<...<l_{M}$) denotes an ordered set of
$M$ different indices from $\{1,2,...,N\}$. For the states presented here, $%
M\leq 2.$

Our procedure is as follows. First we cool the system to the ferromagnetic
ground state $\left\vert \mathbf{0}\right\rangle$, where all spins are down.
Then we encode the state $\left\vert \varphi (0)\right\rangle =\alpha
\left\vert 0_{L}\right\rangle +\beta \left\vert 1_{L}\right\rangle $ at one
end of the chain. The initial state of the whole system is then
$\left\vert \Phi (0)\right\rangle =(\alpha \left\vert 0_{L}\right\rangle
+\beta \left\vert 1_{L}\right\rangle )\otimes \left\vert \mathbf{0}%
\right\rangle$. Note that we are using $\left\vert i_L\right\rangle$ to denote a logical basis state, emphasizing that our physical spins are
\textit{encoded} into a \textit{logical} qubit.

\section{DFS encodings}

We will consider several different encodings with the potential for reliable
information transfer and will provide the best overall solution at the end
of our analysis. In each case we are motivated to consider a state encoded
in a DFS given the universality properties of the states with respect to the
Heisenberg exchange interaction. We will consider two, three, and four-qubit
DFSs which encode one logical qubit using a subsystem of three or four
physical qubits, respectively.

The two-qubit encoding uses $\left\vert 0_L\right\rangle=\left\vert
01\right\rangle$ and $\left\vert 1_L\right\rangle=\left\vert 10\right\rangle$%
. For three-qubit encoding \cite{Knill:99a}, $\left\vert 0_{L}\right\rangle
=\alpha _{0}(\left\vert 010\right\rangle -\left\vert 100\right\rangle )/%
\sqrt{2}+\beta _{0}(\left\vert 011\right\rangle -\left\vert 101\right\rangle
)/\sqrt{2}$, $\left\vert 1_{L}\right\rangle =\alpha _{1}(2\left\vert
001\right\rangle -\left\vert 010\right\rangle -\left\vert 100\right\rangle )/%
\sqrt{6}+\beta _{1}(-2\left\vert 110\right\rangle +\left\vert
011\right\rangle +\left\vert 101\right\rangle )/\sqrt{6}$. The notation $%
\alpha _{0},\beta _{0}$ means that the arbitrary superposition of the state $%
(\left\vert 010\right\rangle -\left\vert 100\right\rangle )/\sqrt{2}$ and
state $(\left\vert 011\right\rangle -\left\vert 101\right\rangle )/\sqrt{2}$
does not change the state $\left\vert 0_{L}\right\rangle $ and $\alpha
_{1},\beta _{1}$ has the same meaning. For the four-qubit encoding \cite%
{Lidar:98}, $\left\vert 0_{L}\right\rangle =(\left\vert 0101\right\rangle +\left\vert 1010\right\rangle -\left\vert 0110\right\rangle -\left\vert
1001\right\rangle )/2$, $\left\vert 1_{L}\right\rangle =(2\left\vert 0011\right\rangle +2\left\vert 1100\right\rangle -\left\vert 0110\right\rangle
-\left\vert 1001\right\rangle -\left\vert 0101\right\rangle -\left\vert 1010\right\rangle )/\sqrt{12}$. The three-qubit and four-qubit DFSs protect a
single logical qubit from collective errors of any type (bit-flip, phase-flip or both) with the three-qubit encoding being the most efficient
\cite{Knill:99a,Lidar:98}.

The fidelity between the received
state and ideal state $\left\vert \varphi (0)\right\rangle $ is defined by $%
F=\sqrt{\left\langle \varphi (0)\right\vert \rho (t)\left\vert \varphi
(0)\right\rangle },$ where $\rho (t)$ is the reduced density matrix at the
receiving end. (We let our ideal final state be represented by the same
vector as our initial state although it is actually at the end, rather than
the beginning, of the chain.)

For example, for the two-qubit encoding, the fidelity at sites $N-1,N$ is $%
F=\left\vert \alpha ^{\ast }A_{N}+\beta ^{\ast }A_{N-1}\right\vert $ where $%
A_{l}=\beta f_{1,l}+\alpha f_{2,l}.$ Similarly we can calculate the fidelity
for the three- and four-qubit encodings, although the expressions are
understandably much more complicated.

From the initial state, $\left\vert \Phi(0)\right\rangle$,
the system undergoes the time evolution given by Eqs.~(\ref{eq:oe})-~(\ref%
{eq:ffe}). The transmission amplitude $f_{j,l}(t)$ describes the propagation and the dispersion characterizes the state transfer. Note also that if the
initial state involves a fixed excitation number, the magnetic field will only produce a global phase which will not affect the fidelity. (See, for
example, the fidelity of the two-spin encoding above.) So for a fixed excitation number, we will neglect the effect of magnetic field be
fixed $h=1.0$. 

\begin{figure}[tbph]
\centering
\includegraphics[scale=0.7,angle=0]{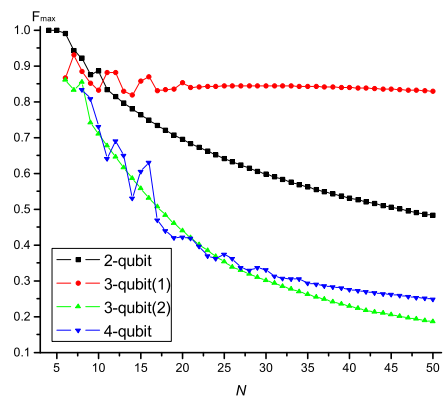}
\caption{Length dependence of the maximum fidelity that can be obtained when transferring a quantum state $(\left\vert 0_{L}\right\rangle +\left\vert
1_{L}\right\rangle )/\protect\sqrt{2}$ from one end of the chain to the other end. } \label{fig:DFSstates}
\end{figure}

\section{DFS-encoding results}

In Fig.~\ref{fig:DFSstates} we compare the maximum fidelity for logical state transfer when $\left\vert \Phi(0)\right\rangle = (\left\vert
0_{L}\right\rangle +\left\vert 1_{L}\right\rangle )/\sqrt{2}$ for different logical/encoded states. Throughout this article the maximum fidelity is
found in the time interval [0,100] since the first peak, which is the maximum, is obtained in this time interval for $N\leq 80$. For the 3-qubit
encoding, we use 3-qubit(1) and 3-qubit(2) to signify one excitation or two excitations in
the chain, for example a state in 3-qubit(1) has $\alpha _{0}=\alpha _{1}=1$%
.

For the two-spin encoding with $N=4,5,$ $F_{\max }\approx 1$ thus near-perfect state transfer can be obtained for these values of $N$. However, for
these and almost all other logical states using two, three, or four spins, the maximum decreases quickly with increasing $N$. There is one exception in
the space of the 3-qubit encoding. For a particular state, the fidelity decreases very slowly with increasing $N$. We will explore this
particular case, which shows impressive variance in the fidelity in Fig.~\ref%
{fig:DFSstates}, and show how it can be used to reliably transfer an
arbitrary qubit state through the chain.

\begin{figure}[tbph]
\centering
\includegraphics[scale=0.6,angle=0]{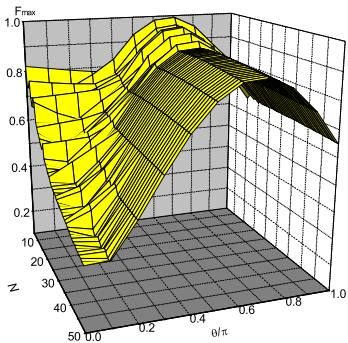}
\caption{The maximum fidelity as a function of $N$ and $\protect\theta$. $%
\protect\phi=0$. The results are for the time interval $[0,100] $.} \label{fig:1}
\end{figure}

\section{Three qubits}

It turns out that two excitations in the chain do not allow for reliable transfer. So we will consider the 3-qubit(1) DFS. In Fig.~\ref{fig:1} we plot
the maximal fidelity, $F_{\max }$, as a function of $N$ and $\theta $ when $\phi =0$ and $t\in \lbrack 0,100]$ for an arbitrary initial state $\cos
(\frac{\theta }{2})\left\vert 0_{L}\right\rangle+\sin (\frac{\theta }{2} )e^{i\phi }\left\vert 1_{L}\right\rangle $.
For $N$ from $6$ to $50$, a maximum is achieved at $\theta =2\pi /3$. And
surprisingly, $F_{\max}$ decreases very slowly with increasing $N$ for a
wide range of $\theta $. For example, in the range $\theta \in \lbrack
0.5\pi ,0.8\pi ]$, for any site $N\leq 50$, $F_{max}>0.8$. So if the state
is encoded in this range, the fidelity is exceptionally large for a quite
long chain. In fact, we have found that $F\approx 0.7$ \textit{after
traversing a spin chain of two hundred spins! } Therefore, \textit{we can
achieve a very reliable state transfer since the fidelity is large and
provides a significant robustness to errors in the initial encoding, or
during transport,} since a variation in the encoded state does not
significantly affect the long distance trend in the fidelity. It still
decreases very slowly over long distances. We next show how to take
advantage of this remarkable state.

\begin{figure}[tbph]
\centering
\includegraphics[scale=0.7,angle=0]{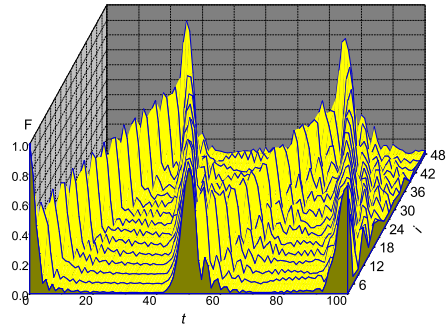}
\caption{The time evolution of the fidelity at various sites $i$ for a long chain ($N$=48). We use the fidelity at sites $i$ to denote that we receive
the state at sites $i-2,i-1,i$. The initial state is $1/\protect\sqrt{2}%
[\left\vert 001\right\rangle -\left\vert 100\right\rangle ]\otimes\left\vert \mathbf{0}\right\rangle$.} \label{fig:2}
\end{figure}

\section{Efficient encoding for encoded qubits}

When $\phi =0,\theta =2\pi /3$ the 3-qubit(1) encoding can be written as $%
\left\vert \Psi \right\rangle =\frac{1}{\sqrt{2}}(\left\vert
001\right\rangle -\left\vert 100\right\rangle )=\frac{1}{\sqrt{2}}%
(\left\vert 01\right\rangle _{13}-\left\vert 10\right\rangle _{13})\otimes \left\vert 0\right\rangle _{2}$, i.e. the first and third spins are in a
singlet state. In order to show the high-fidelity transfer of this state, in Fig.~\ref{fig:2} we plot the time evolution of the fidelity at every site
of a chain of length $N=48$. At $t=0$ the initial state is encoded at the sites of the first and third spins, then it evolves freely under the
Heisenberg Hamiltonian. The time dependence is given by $\left\vert \Phi (t)\right\rangle =1/\sqrt{2}[\sum_{i=1}^{N}(f_{3,i}-f_{1,i})c_{i}^{\dag
}]\left\vert \mathbf{0}\right\rangle $. At time $t$, the fidelity at the $%
i-2,i-1,i$ sites in the interior of the chain is $F\approx 0.5$ implying
only partial information is located at these sites. However, at the end of
the chain ($i=48$), the fidelity shows a peak ($F_{1}\approx 0.86,t\approx 25
$). After this, the wave is reflected by the boundary, and starts to
propagate back. This behavior can be interpreted as a wave which broadens
inside the chain, but when arriving at the boundary it becomes narrower
which enhances the fidelity. Thus we have an end-effect of the chain. From
Fig.~\ref{fig:2}, the oscillation of the state between boundaries is $%
T\approx 50$ and at time $t_{k}\approx 25+(k-1)T,$ a maximum is achieved at the end
of the chain $N=48$, where $k$ denotes the k$th$ peak. For example, at time $%
t=75$, the state will travel to the other end once more (the second peak $%
F_{2}\approx 0.76$), but $F_{2}<F_{1}$ which shows some reduction of fidelity with each pass, but still relatively high.

We have shown that the state $1/\sqrt{2}[\left\vert 001\right\rangle -\left\vert 100\right\rangle ]$ can be transferred through the chain with high
fidelity even when the chain is quite long. For quantum communication, we need to transfer an \textit{arbitrary} state with high fidelity. In this case
it important to realize that an encoding which uses: $\left\vert 0_{L}\right\rangle =\left\vert 000\right\rangle $ and $\left\vert 1_{L}\right\rangle
=1/\sqrt{2}[\left\vert 001\right\rangle -\left\vert 100\right\rangle ]$ can reliably transfer the state since the vacuum state is fixed throughout.
Using this encoding we can fully utilize the extremely reliable state $\left\vert \Psi\right\rangle_L = \alpha \left\vert 000\right\rangle +\beta
/\sqrt{2}[\left\vert 001\right\rangle -\left\vert 100\right\rangle ]$.


\begin{figure}[tbph]
\centering
\includegraphics[scale=0.6,angle=0]{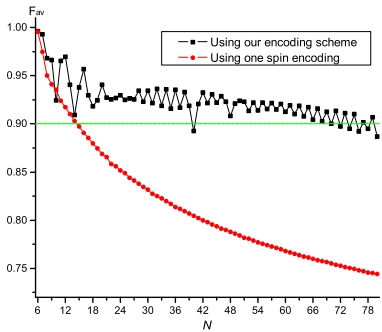}
\caption{Comparison of the average fidelity that can be obtained as a function of chain length $N$ for two different encoding schemes. This has been
optimized with respect to time and magnetic field.} \label{fig:logicalstate}
\end{figure}

In Fig.~\ref{fig:logicalstate} we compare two different encoding schemes for state transfer, circles for the original single-spin encoding
\cite{Bose:03} and squares for ours. The average fidelity $\frac{1}{4\pi }\int \left\langle \varphi _{in}\right\vert \rho _{N}(t)\left\vert \varphi
_{in}\right\rangle d\Omega$ is
\begin{eqnarray}
F_{av} &=&\frac{1}{3}+\frac{1}{3}\mbox{Re}[\frac{G_{N}-G_{N-2}}{\sqrt{2}}]+%
\frac{1}{3}\left\vert \frac{G_{N}-G_{N-2}}{\sqrt{2}}\right\vert ^{2}  \notag
\\
&&+\frac{1}{6}[1-\left\vert G_{N}\right\vert ^{2}-\left\vert G_{N-1}\right\vert ^{2}-\left\vert G_{N-2}\right\vert ^{2}].
\end{eqnarray}%
where $G_{i}=\frac{1}{\sqrt{2}}$ $[f_{3,i}(t)-f_{1,i}(t)]$. The results are maximized over the time
interval [0,100] and magnetic field $h\in [0,2]$. (Unlike transferring the $%
\left\vert 1_{L}\right\rangle$ state, here the magnetic field can be
adjusted to enhance the fidelity since the phase is now not negligible.) For
single-spin encoding, $F_{av}$ decreases relatively quickly with increasing $%
N$. However, \textit{using our scheme, $F_{av}$ decreases very slowly with
increasing $N$ and $F_{av}$ is always relatively high even in a long chain ($%
N\gtrsim 80$). For example $N\approx 70,$ $F_{av}$ is still greater than}
0.9.

\section{Protocols for improving reliability}

Here we present two protocols for increasing the reliability of state
transfer using our encoded state. One is due to Giovannetti et al. \cite%
{Giovan:06,Shizume:07} and the other is based on, but a generalization of, a
protocol by Burgarth and Bose \cite{Burgarth/Bose:05}.

In the first protocol we consider, Giovannetti et al. \cite%
{Giovan:06,Shizume:07} showed that the reliability of the the one-spin encoding can be enhanced using a memory. In this protocol, the receiver swaps the
state at the end of the chain to a quantum memory for decoding at a later time. This process is repeated for later times, with each swap and storage
increasing our overall chance of success. Fig.~\ref{fig:2} shows the variation of the fidelity from which we may infer the chance of success
for our protocol. If we perform the swap operation as in Ref.~\cite%
{Giovan:06} at $t=25$, the probability that the $\left\vert 1_{L}\right\rangle $ state has been swapped to the first memory is $\eta=\left\vert
_{46,47,48}\left\langle 1_{L}\right\vert \left\langle \mathbf{0}\right\vert e^{-iHt_{1}}\left\vert 1_{L}\right\rangle _{1,2,3}\left\vert
\mathbf{0}\right\rangle \right\vert ^{2}\approx0.86^{2}$, which corresponds to the square of the first peak value at $N=48$. Performing additional swap
operations at some later optimal time will increase our already large probability for success, just as it does in the original protocol for the
single-spin encoding. 

We next provide a protocol which can confirm if a state was indeed transferred appropriately, which is a generalization of that presented in
Ref.~\cite{Burgarth/Bose:05}. We begin with two spin chains which are initially decoupled and proceed as follows. First the logical state is encoded
into the first and third site of the first chain. Then a logical $X$ gate is performed on spins one and three of the second chain conditioned on the
logical state of the first chain being zero \cite{CNOTnote} (using the standard ordered basis). The form of the logical X is
\begin{equation*}
X_L^{(1,3)} = \left(%
\begin{array}{cccc}
0 & 1 & 0 & 0 \\
1/\sqrt{2} & 0 & 0 & 1/\sqrt{2} \\
-1/\sqrt{2} & 0 & 0 & 1/\sqrt{2} \\
0 & 0 & 1 & 0%
\end{array}%
\right).
\end{equation*}
(From this it is straight-forward to obtain the logical CNOT which is also
to be performed on spins one and three.) The state of the chain is then $%
\left\vert \Psi\right\rangle = \alpha\left\vert
0_L\right\rangle\otimes\left\vert 1_L\right\rangle + \beta\left\vert
1_L\right\rangle\otimes\left\vert 0_L\right\rangle$ where the first factor
for the first chain and the second for the second chain. We now let both
chains evolve freely. After the same amount of time as above, we perform a
logical CNOT operation to decode the operation. Performing a measurement on
the last three spins of the second chain and finding $\left\vert
000\right\rangle$ will confirm that the state has been reliably sent through
the chain. This confirmation, along with the very high fidelity of our
protocol, provides a high probability of reliable transfer, along with
confirmation.


\section{Conclusions}

We have presented several results for a decoherence-free/noiseless subspace
encoded qubit which is transferred through an unmodulated spin chain.
Although many of the encoded states were not more reliable than the
single-spin encoding, we have found remarkable results. A most striking
result is that when the initial state is a 3-qubit(1) state, $\cos (\frac{%
\theta }{2})\left\vert  0_{L}\right\rangle +\sin (\frac{\theta }{2})e^{i\phi }\left\vert  1_{L}\right\rangle $, with $\theta$ near $2\pi /3$ and $\phi$
near zero, is sent through the spin chain, the fidelity is incredibly high. Surpassing any known result so far. Even out to two hundred spins, the
fidelity is quite high($\approx0.7$). For transferring an arbitrary state, we have found a very high fidelity based on these results. For example, when
$N \leq 70$ the fidelity $F\geq 0.9$. This is a remarkable result for a simple two-spin encoded state and experimentally viable due to its
Heisenberg-mediated transfer with control assumed only on two of the three at the ends of spins of the chain.

Furthermore, we have shown that our protocol can be combined with a protocol
using a local memory to enhance the fidelity beyond an already impressive
value. We have also presented a protocol for confirmation of the receipt of
the state at the other end. We therefore believe this is by far the best
protocol to date for the transfer of a quantum state through an unmodulated
spin chain.


\section*{Acknowledgments}

This material is based upon work supported by NSF-Grant No. 0545798 to MSB.
ZMW thanks the China Scholarship Council. We acknowledge C. A. Bishop and
Y.-C. Ou for helpful discussions.



\begin{thebibliography}{18}
\bibitem[CNO()]{CNOTnote}
\bibinfo{note}{{This is equivalent to performing an $X_L$ gate on the first
  state of the first chain, followed by a logical CNOT gate on the first spin
  of the first chain, which has as its target the first spin of the second
  chain, followed by the an $X_L$ gate on the first spin of the first chain. In
  general a logical operation can be determined by using the unitary which
  changes basis, and calculating the logical operation from the standard one.
  See for example \cite{Byrd/etal:05}.}}
\end{thebibliography}


\end{document}